\documentclass[prc,onecolumn,tightenlines,12pt]{revtex4}
\usepackage{graphicx}
\usepackage{dcolumn}
\usepackage{amsmath}
\begin{document}

\title{Randomly incomplete spectra and intermediate statistics}

\author{O. Bohigas$^{1}$ and M. P. Pato$^{1,2}$}

\affiliation{
$^{1}$CNRS, Universit\'e Paris-Sud, UMR8626, \\
LPTMS, Orsay Cedex, F-91405, France \\
$^{2}$Inst\'{\i}tuto de F\'{\i}sica, Universidade de S\~{a}o Paulo\\
Caixa Postal 66318, 05315-970 S\~{a}o Paulo, S.P., Brazil}

\begin{abstract}
By randomly removing a fraction of levels from a given spectrum a model is
constructed that describes a crossover from this spectrum to a Poisson
spectrum. The formalism is applied to the transitions towards Poisson from 
random matrix theory (RMT) spectra and picket fence spectra. It is
shown that the Fredholm determinant formalism of RMT extends naturally 
to describe incomplete RMT spectra.   
\end{abstract}
\maketitle

It is by now well established that, in the semiclassical limit, 
spectral properties of physical systems
whose underlying classical motion is chaotic are consistent with the 
predictions of the Wigner-Dyson random matrix theory\cite{Boh82},
while those with an underlying regular motion behave as an uncorrelated 
sequence of levels (Poisson statistics)\cite{Berry} (see \cite{Guhr}
for a review). Matrix models have been proposed to describe
transitions between these two extreme behaviors. In  these
interpolating matrix models the transition is generated by 
reducing the magnitude of the off-diagonal elements with respect
to the diagonal ones. This can be achieved in different ways, for instance
by decreasing the variance of all off-diagonal elements\cite{Pato}, by 
making the matrices banded\cite{Casati} or by introducing a power-law 
decay\cite{Mirlin} or more complicated schemes\cite{Moshe}. In this kind
of models the invariance of the ensemble probability distribution
under unitary transformation is broken. Spectral distributions of 
some invariant ensembles may also show deviation from the Wigner-Dyson 
statistics but without reaching the Poisson limit\cite{Mutt,Bogom}. 
Apparently symmetry breaking is necessary to completely
eliminate the correlations among levels\cite{Krav}.

Despite that classical analogy does not support the assumption of a
unique crossover from chaotic to regular regimes,
the existence of a third universal statistics has been conjectured
motivated by the Anderson model description of the metal-insulator 
transition (MIT)\cite{Aronov}. In the MIT, localized states of the
insulator phase obey Poisson statistics while the extended states of 
the metal phase show a Wigner-Dyson behavior. The third universal 
behavior would correspond to statistics observed at the 
critical point. The main characteristics of this 
behavior would be the multifractality of the eigenstates, nearest 
neighbor spacing distribution (NND) exhibiting 
a linear level repulsion with a slope at the origin steeper than the Wigner 
case and with an exponential decay for large separations in contrast to 
the Gaussian standard decay. Long range statistics like number variance 
would increase linearly as in the Poisson case 
but with a smaller slope\cite{Garcia}. Although ensembles have 
been found, invariant or not, that show some of these 
properties, the extent of their validity has not yet been 
established.

Perhaps the simplest model with these spectral features is the short 
range plasma model\cite{Bogom1}. It consists in a 1D Coulomb gas 
model for which the range of the interaction is restricted to
a finite number of neighboring levels. An appealing feature 
of the model is its amenability to analytical treatment. It was 
found that the intermediate statistics obtained when only adjacent 
pairs interact, denoted by the authors as the semi-Poisson statistics, 
gives an approximately good description of spectral properties of 
some diffractive billiards\cite{Bogom2}.

It has been shown that semi-Poisson statistics can also be obtained 
in a completely independent way as a particular case of a family of
statistics termed by their authors the daisy models \cite{Flores}. 
It consists in removing every other $r$ levels from an uncorrelated 
spectrum. The particular case $r=1$ (every other level) corresponds 
exactly to the semi-Poisson statistics. Notice also that it has been 
known for quite a long time that spectral properties of two different 
classes of spectra can be related by the operation of dropping levels 
from one of them, namely measures of the symplectic 
and the orthogonal ensembles of RMT\cite{Mehta} can be connected in 
this way. 

In the above cases, levels are removed in a correlated way 
(every other level) and the operation results in a more correlated 
spectrum (from GOE to GSE, from Poisson to semi-Poisson). Our purpose 
here is to investigate the same operation but performed randomly. 
Specifically, we consider an infinite spectrum and, after dropping 
at random a fraction $1-f$ of levels, the remaining fraction $f$ is 
studied ($0<f<1$). In order to keep to unity the average level density
the remaining spectrum is correspondingly contracted. 
To determine the statistical properties of the transformed
spectra we resort to the formalism developed in our previous 
work\cite{Boh} in which the problem of randomly incomplete 
spectra was considered. We showed that when $f\rightarrow 0,$ 
irrespective of the nature of the initial spectrum, the statistical 
properties approach those of an uncorrelated spectrum 
(the Poissonian spectrum is a fixed point of the operation). 
Therefore, the random dropping operation 
generates models whose statistics are intermediate between those of 
the initial spectrum and Poisson statistics. Obviously it can also 
be seen that the statistical properties of a Poisson
sequence are not affected by this dropping operation.  

In the present communication we discuss properties of models constructed 
starting with RMT spectra and with a picket fence of equally spaced 
levels.  We show that in the first case a family of intermediate 
statistics parameterized by the fraction $f$ of the remaining levels 
is generated that shows features similar to those of the intermediate 
critical statistics. In particular, the $f=1/2$ case is compared 
with the semi-Poisson statistics. In the picket fence model we show that 
the $f=1/2$ member of the family reproduces the spectral properties of 
a sequence of levels weakly confined by a log-normal potential (see below). 

For later use, we recall some notations and results of
\cite{Boh}. We consider a spectrum $\rho(E)=\sum\delta(E-E_i)$ with
mean spacing $<\rho>=1.$ The two-point cluster function 
$Y_2(x)=1-<\rho(E-x/2)\rho(E+x/2)>$ gives the disconnected part
of the two-point correlation function. One has the basic relation 

\begin{equation}
\hat{y}_2\left(x\right)=Y_2\left(\frac{x}{f}\right) \label{60}
\end{equation}
that expresses the two-point cluster function of a spectrum with a 
fraction $1-f$ of missing levels in terms of the same function of 
the complete spectrum. We use capital letters to
denote the quantities of the complete intial spectrum and small cases
with a superscript for the incomplete ones. Similar scaling relations
hold for higher-cluster functions. From (\ref{60}) other statistical 
measures can be easily derived. For instance the form factor 
$K\left(\tau\right)=1-B\left(\tau\right),$ where $B(\tau)$ is the Fourier 
transform of the cluster function, transforms as

\begin{equation}
\hat{k}\left(\tau\right)=1-f+fK\left(f\tau\right) .  
\end{equation}
Similarly, the number variance $\hat{\sigma} ^{2}$ (variance of the number 
of levels contained in an interval of length $L$) of the 
transformed spectrum is expressed in terms of the same quantity of
the complete spectrum 

\begin{equation}
\hat{\sigma} ^{2}\left( L\right) =\left( 1-f\right) L+f^{2}\Sigma ^{2}
\left( \frac{L}{f}\right).  \label{64}
\end{equation}
The important feature of this relation is the appearance
of a linear term suggesting the same behavior as for critical
statistics. In particular, the Poisson expression is recovered 
when $f\rightarrow 0$.

Another set of statistical measures are the $E(n,s)$ functions ($n$-level
probability functions) that give the
probabilities of finding $n$ levels ($n=0,1,2...,$) inside a segment 
of length $s.$ If their expressions $E(n,s)$ for the complete spectrum 
are known then the first ($n=0$) of these functions (gap probability 
function), when only a partial fraction $f$ of levels taken at random 
remains,is given by

\begin{equation}
\hat{e}(s,f)=\sum_{k=0}^{\infty }\left( 1-f\right)^{k}E(k,\frac{s}{f}), \label{4}
\end{equation}
which follows from the fact that $1-f$ is the probability that one level was 
dropped. By the same argument the NND $\hat{p}(s,f)$ is given by

\begin{equation}
\hat{p}(s,f)=\sum_{k=0}^{\infty }\left( 1-f\right) ^{k}P(k,\frac{s}{f}), \label{28}
\end{equation}
where the $P(k,s)$ are the density distributions of the spacings between
two levels containing $k$ levels inside the complete sequence
($P(0,s)$ is the NND of the complete sequence, $\hat{p}(s,f)$ the one
corresponding to the incomplete sequence). This expression was first proposed 
as an $\it {anzatz}$\cite{Bilpuch} and  in Ref. \cite{Mitchell} it 
is shown that the coefficients $f\left( 1-f\right) ^{k}$ 
maximizes Shannon entropy with constraints appropriately defined.  

The above equations show that these expressions for the gap probability
and NND of the transformed spectrum are the generating functions of 
all the $n$-level probability functions and spacings distributions of the 
complete spectrum. Indeed, by defining the generating function   

\begin{equation}
G(t,z)=\sum_{k=0}^{\infty }\left(-1\right)^{k}\left(z-1\right)^{k}E
\left(k,t\right)
\end{equation}
such that 

\begin{equation}
E(n,t)=\frac{\left(-1\right)^{n}}{n!}\left[\frac{\partial G(t,z)}
{\partial z^{n}}\right]_{z=1}
\end{equation}
then, from Eq. (\ref{4}), the identification
$\hat{e}(s,f)=G(\frac{s}{f},f)$ can be made. Of course,  an equivalent 
identification also holds for the incomplete NND  given 
by Eq. (\ref{28}).

As a trivial example of the above relation, the expressions 
$E(n,s)=\frac{s^n}{n!}\exp(-s)$ for the Poissonian $n$-probability 
functions are generated by the function $G(t,z)=\exp(-tz).$ 
Taking $t=\frac{s}{f}$ and $z=f$ we can check that they are not 
affected by the dropping level operation.  

We apply now the above formalism to initial spectra of standard RMT. 
By observing that the $\hat{e}$-function is the generating 
function of the $n$-level probability functions we can establish 
connections between the statistical properties of the 
transformed spectra and the RMT Fredholm determinants. These are 
the determinants $D(t,z)=\det \left(1-zK\right)$ and 
$D_{\pm}(t,z)=\det \left(1-zK_{\pm}\right)$ where $K$ and $K_{\pm}$
are, respectively, the integral  operators with kernels 
$K(x,y)=\frac{1}{\pi}\sin(x-y)/(x-y)$ and 
$K_{\pm}(x,y)=K(x,y)\pm K(x,-y)$ defined 
on $L^2 ([0,\pi s])$\cite{Mehta}. 
 
Starting with the unitary Gaussian ensemble (GUE), the $n$-probability
functions (in the sequel, the index $\beta$ with $\beta=1,2$ and $4$ 
denotes quantities of the orthogonal, unitary and symplectic 
ensembles, respectively) are given in terms of $D(t,z)$ 
by

\begin{equation}
E_2(n,t)=\frac{\left(-1\right)^{n}}{n!}\left[\frac{\partial D(t,z)}
{\partial z^{n}}\right]_{z=1} \label{19}
\end{equation}
For GOE the even and the odd  $n$-probability functions are given by

\begin{equation}
E_{1}(2n,t)=\sum_{k=0}^{n} E_{+}(k,t) -\sum_{k=0}^{n-1} E_{-}(k,t) 
\end{equation}
with $E_{-}(-1,t)=0$ and

\begin{equation}
E_{1}(2n+1,t)=\sum_{k=0}^{n}\left[E_{-}(k,t) - E_{+}(k,t)\right],
\end{equation}
where $E_{\pm}$ are expressed in terms of $D_{\pm}$ as $E_{2}$ 
and $D$ in (\ref{19}).
In terms of $E_{\pm},$ the symplectic $n$-probability functions
are given by 

\begin{equation}
E_{4}(2n,t)=\frac{1}{2}\left[E_{+}(k,2t)+ E_{-}(k,2t)\right] 
\end{equation}
On the other hand Jimbo {\it et al.}\cite{Jimbo} have shown that 
the determinant $D(t,z)$ is given by

\begin{equation}
\ln D(t,z)=\int^{\pi t}_{0}\frac{\sigma(x,z)}{x}dx ,
\end{equation}
where $\sigma(x,z)$ is the solution of the differential equation 

\begin{equation}
\left(x\frac{d^{2}\sigma}{dx^{2}}\right)+4\left(x\frac{d\sigma}{dx}
-\sigma\right)\left[x\frac{d\sigma}{dx}-\sigma+\left(\frac{d\sigma}
{dx}\right)\right]=0
\end{equation}
that satisfies the boundary condition $\sigma(x,z)\sim-\frac{z}
{\pi}x$ when $x\rightarrow0$. It has also been shown that
$D_{\pm}(t,z)$ are given in terms of $D(t,z)$ as

\begin{equation}
\ln D_{\pm}(t,z)=\frac{1}{2}\ln D(t,z) \pm \frac{1}{2}\int^{t}_{0}
dx\sqrt{-\frac{d^2}{x^{2}}\ln D(t,z)}
\end{equation}

From the above we derive that the incomplete gap functions
in the three cases are connected with the respective determinants
$D(t,z)$ and $D_{\pm}(t,z)$ by the relations

\begin{equation}
\hat{e}_2(s,f)=D(\frac{s}{f},f)
\end{equation}
for the unitary,

\begin{equation}
\hat{e}_{1}(\frac{s}{f},f)=\frac{1}{2-f}\left[ D_{+}(\frac{s}{f},
2f-f^2)+(1-f)D_{-}(\frac{s}{f},2f-f^2)\right]
\end{equation}
for the orthogonal, after some algebra, and finally for the symplectic

\begin{equation}
\hat{e}_{4}(\frac{s}{f},f)=\frac{1}{2}\left[ D_{+}(\frac{2s}{f},f)+
D_{-}(\frac{2s}{f},f)\right].
\end{equation}

The above equations provide exact expressions for the gap 
functions of both complete and randomly incomplete RMT spectra.
They are one of the main results of the present communication as 
they provide a physical interpretation for all the real values 
of the parameter $z$ in the interval $[0,1]$ appearing in the 
Fredholm determinant.  With  $z=1$ they have been used to derive 
asymptotics of the spacings for large separations\cite{Basor}. 
We now show that when spectra are incomplete a major change 
in their asymptotics appears. This follows from the fact that 
with $0<z<1$, the function $\sigma(x,z)$  behaves, 
when $x\rightarrow\infty,$ as\cite{Tang}

\begin{equation}
\sigma(x,z)=\frac{x}{\pi}\ln(1-z)
\end{equation}
Substituting into the above equations, one finds that 
to leading order 

\begin{equation}
\hat{e}_{\beta}(s,f)=\exp\left[\frac{s}{f}\ln(1-f)\right] \label{15}
\end{equation}
for $\beta=1,2$ while $f$ has to be multiplied by $2$ for $\beta=1$. 
When $f\rightarrow0$ one gets the Poisson behavior $\exp(-s)$
irrespective of the value of $\beta.$

Let us  remark that although we discuss here the $\hat{e}$-function, 
the identification of the incomplete NND $\hat{p}$-function as a 
generating function of spacing distributions can also be used 
to perform a similar analysis\cite{For}. 

We can now compare the incomplete GOE case with the semi-Poisson 
model. This model gives rise to a linear number variance with 
slope $1/2$. By taking $f=1/2$ in (\ref{64}), the incomplete 
GOE spectrum has the same behavior for large values of $L$  apart 
from a small contribution of the $\Sigma^2$ term. Considering the 
form factor we have for the incomplete sequence

\begin{equation}
\hat{k}(\tau)=1-f+f\left\{ 
\begin{array}{rl}
2f\tau+f\tau\ln(1+2f\tau),& \  0\leq f\tau \leq 1 \\ 
2-f\tau\ln\left(\frac{2f\tau+1}{2f\tau-1}\right), &\ 
f\tau\geq 1
\end{array}
\right.  \label{17}
\end{equation}
and for the semi-Poisson model

\begin{equation}
K(\tau)=\frac {2+\pi^{2}\tau^{2}}{4+\pi^{2}\tau^{2}}.
\end{equation}
In Fig 1a these two functions are compared. They do not coincide
but they show strong similarities: they both start with a value $1/2$
at the origin, reflecting an identical (lack of) rigidity and for 
large $\tau$ they both tend to $1$ like $1/\tau^{2}.$ Concerning 
the spacing distributions, recall first that the semi-Poisson 
NND is given by $4s\exp(-2s)$\cite{Bogom1}. Close to the origin only 
the $k=0$ term contributes in (\ref{28}), therefore  
$\hat{p}(s,f)\sim P_{GOE}(0,\frac{s}{f})\sim \frac{s}{f}\frac{\pi^2}{6}$ 
leading with $f=1/2$ to a slightly smaller slope for the 
incomplete spectrum. On the other hand, taking $f=1/2$ 
in (\ref{15}) we have for large separations a decay 
$\exp\left[-2(\ln2)s\right]$ slower than the semi-Poisson one. 
This comparison is illustrated in Fig 1b.
In summary, although with similarities, the incomplete GOE model 
presents differences with respect to the semi-Poisson model. 

It is worth noticing that although our procedure interpolates 
between initial and final (Poisson) spectra, it gives rise to 
different results from the ones resulting
from superposing in an uncorrelated way different spectra\cite{Mehta}. 
In this latter case, for example, level repulsion is destroyed while 
the dropping mechanism presented here preserves it for
all values of $f$.     

Let us now apply the dropping procedure to a picket fence spectrum 
defined as a sequence of points located, say, 
at $...-3/2,-1/2,1/2,3/2...$. From the definition, one can write  

\begin{equation}
P(n,s)=\delta \left( s-n-1\right)  \label{84}
\end{equation}
and 

\begin{equation}
E(n,s)=\left\{ 
\begin{array}{rl}
1-\left| s-n\right| ,& \ \left| s-n\right| \leq 1 \\ 
0, & \  \left| s-n\right| \geq 1
\end{array}
\right.  \label{85}
\end{equation}
and for the two-point cluster function

\begin{equation}
Y_{2}\left( x\right) =1-\sum_{n=0}^{\infty
}\delta \left[x-(n+1)\right]  \label{90}
\end{equation}
which follows from (\ref{84}) and the general relation 
$Y_2(x)=1-\sum_{n=0}^{\infty}P(n,x).$   
The picket fence is the most 'correlated' spectrum and its 
rigidity reflects, for instance, in the smallness of the 
number variance 

\begin{equation}
\Sigma ^{2}\left( L\right) =L-[L]-\left(L-[L]\right)^2
\end{equation}
where $[L]$ stands for the integer part of $L$. 

For later comparison with the behavior of weakly confined 
eigenvalues, consider now what we denote as the 'continuous' version 
of the picket fence spectrum, namely each point is randomly, 
independently and uniformly distributed inside an interval of unit 
length around the values $...-3/2,-1/2,1/2,3/2,...$. In this case, 
the $\delta$-functions in  (\ref{84}) of the spacing distributions 
become the ``triangles''

\begin{equation}
P(n,s)=\left\{ 
\begin{array}{rl}
0, & s\leq n \\ 
s-n, & n\leq s\leq n+1 \\ 
s-n-2, &  n+1\leq s\leq n+2 \\ 
0, & n+2\leq s
\end{array}
\right. \label{88}
\end{equation}
that preserve the normalization $<1>=1$ and $<s>=n+1$. 
From (\ref{88}) we find that the cluster function takes the simple 
expression

\begin{equation}
Y_{2}(x)=\left\{ 
\begin{array}{rl}
1-x, & 0\leq x\leq 1 \\  
0, & 1\leq x
\end{array}
\right.
\end{equation}
from which follows

\begin{equation}
K(\tau)=1-\left[\frac{\sin(\pi\tau)} {\pi\tau}\right]^{2}
\end{equation}
and  

\begin{equation}
\Sigma ^{2}(L)=\left\{ 
\begin{array}{rl}
L-L^2 +L^{3}/3, & 0\leq L\leq 1 \\  
1/3, & 1\leq L . 
\end{array}
\right.
\end{equation}
Using the general relations for incomplete spectra we find for
the transformed cluster function

\begin{equation}
\hat{y}_{2}(x)=\left\{ 
\begin{array}{rl}
1-x/f, & 0\leq x\leq f \\  
0, & f\leq x ,    
\end{array}
\right.     \label{105}
\end{equation}
form factor

\begin{equation}
\hat{k}(\tau)=1-f\left[\frac{\sin(f\pi\tau)} {f\pi\tau}\right]^{2}
\end{equation}
and number variance

\begin{equation}
\hat{\sigma} ^{2}(L)=(1-f)L+f^{2}\left\{ 
\begin{array}{rl}
L/f-(L/f)^2 +(L/f)^{3}/3, & 0\leq L\leq f \\  
1/3, & f\leq L .
\end{array}
\right. 
\end{equation}
Finally, substituting (\ref{88}) into (\ref{28}), the NND for 
the incomplete sequence becomes

\begin{equation}
\hat{p}(s,f)=\left\{ 
\begin{array}{rl}
f^{-1}s, &  0\leq s\leq f \\  
(1-f)^{n-1}\left( 1+nf-s\right), & nf\leq s\leq \left( n+1\right) f, 
\ n=1,2,3....
\end{array}
\right.  \label{96}
\end{equation}
(see Fig 2, for illustration).

Let us compare these results with the behavior of the eigenvalues of 
matrix ensembles in which the eigenvalues are confined by a
weak log-normal potential. By that is meant eigenvalues of $N$x$N$ random 
matrices $M$ distributed according to

\begin{equation}
P(M) =\exp\left[-TrV(M)\right] 
\end{equation}
with $V(x)=\ln^{2}\left|x\right|/\gamma$.
It has been shown in Ref. \cite{Bogom} that in 
the limit of weak confinement $\gamma \rightarrow \infty $, the
spectrum, whose $N$ levels tend to locate around the sites of 
a crystal lattice, has,  after unfolding, the following structure. 
Construct $2N$ intervals of length $1/2.$ symmetrically with respect 
to the origin.The  $N$ levels occupy randomly these intervals as 
follows: i) in an interval there is at most one level, ii) intervals 
symmetric with respect to the origin can not be simultaneously 
occupied. This last property introduces long range correlations and 
lack of stationarity (translation invariance). By performing an 
average over the spectrum or, equivalently, by considering only the 
first (or second) half of the spectrum, the effect of this long 
range correlations is washed out and one obtains, for the 
two-point cluster function

\begin{equation}
Y_{2}(x)=\left\{ 
\begin{array}{rl}
1-2x, & 0\leq x\leq 1/2 \\  
0, & 1/2\leq x
\end{array}
\right. \label{110}
\end{equation}
and for the NND  (see \cite{Bogom})

\begin{equation}
P(s)=\left\{ 
\begin{array}{rl}
2s, &  0\leq s\leq 1/2 \\  
2^{-n+1}\left( 1+n/2-s\right), & n/2\leq s\leq \left( n+1\right)/2,
\ n=1,2,3....
\end{array}\right.  \label{102}
\end{equation}
Obviously,  (\ref{110}) and  (\ref{102}) are just  (\ref{105}) and 
(\ref{96}) with $f=1/2,$ corresponding to the continuous picket fence 
for which half of the levels have been randomly removed (see Fig. 2). 
  
Let us finally mention an example of a different sort of mechanism than
dropping or removing points. Here points are located on a line in the
plane and, as a parameter is varied, a fraction of them leave the
line. Specifically, consider the roots of a random polynomial of
degree $N$

\begin{equation}
P_N(z)=\sum^{N}_{k=0} a_k z^k
\end{equation}
where the $a_k$'s are independent Gaussian complex random variables 
(real and imaginary parts centered at zero and variance
$\sigma^2$)\cite{Leb}. If one imposes the symmetry (self-inversive (SI)
symmetry, also called selfreciprocal or conjugate reciprocal)

\begin{equation}
a_{N-k} =\bar{a}_k ,
\end{equation}
where the bar denotes complex conjugate, one can see that the roots of
$P_N(z)$ lie either on the unit circle $C$ or appear in pairs
symmetrically located under inversion with respect to it. The relevant
parameter of the model is $\epsilon=\sigma\sqrt{N}.$ As $\epsilon$
increases, some roots leave $C$ and in the 
limit $\epsilon\rightarrow \infty$, on the average, a 
fraction $\phi=1/\sqrt{3}$ of 
the roots remains on it \cite{Leb,Dun}. On the other hand, in
Ref. \cite{Farmer}, the
restricted class of SI polynomials having all the roots on $C$ has
been considered. It has been found that their statistical properties
coincide with those of eigenvalues of the orthogonal ensemble (OE) 
of random matrices ($\beta=1$). One may then ask wether properties 
of the unrestricted class of SI polynomials, in the 
limit  $\epsilon\rightarrow \infty,$ share some properties with 
the ones corresponding to dropping at random a fraction $1-\phi$ 
of zeros of the restricted polynomials. Consider, for instance, the
NND. In both cases it starts linearly at the origin, with a slope
$\pi^{2}/6$ for the restricted case and  $\pi^{2}/(10\sqrt{3})$ for the
unrestricted case\cite{Leb}, which is five times smaller than what
would result from randomly dropping a fraction $1-\phi$ of points from
an OE sequence, namely $\pi^{2}(6\phi)^{-1}$ (see Eq. (\ref{28})). 
Leaving  $C$ has not the same effect as dropping randomly points on
it. The points who move and locate on the complex plane ``interact''
with those remaining on $C.$  

In conclusion, by dropping at random a fraction of levels of a given 
spectrum a family of spectra is generated. Its statistical
properties are intermediate between those of the initial one and a Poisson
spectrum. Applied to eigenvalues of RMT, the family contains as a 
particular case a model exhibiting some of the features of critical 
intermediate statistics. Fredholm determinants of argument $z$ are one
of the basic structures appearing in RMT (one is usully interested in 
properties corresponding to $z=1$). We show, for $z< 1,$ that
properties of Fredholm determinants characterize also the behavior of
incomplete spectra and that $z$ corresponds to the remaining fraction
of levels. When the dropping procedure is applied to a picket fence 
spectrum we show that the generated family has the statistical
properties of an ensemble of eigenvalues weakly confined. Finally 
we compare with a system for which, when a parameter is varied, a 
fraction of points on a line are not dropped but move on the complex 
plane.

One of us (OB) thanks P. Forrester for an illuminating discussion.
This work is supported in part by the Brazilian agencies CNPq and FAPESP.

{\bf Figure Captions}

Fig. 1 (a) form factor; (b) nearest neighbour spacing distribution (NND). Full
line: incomplete ($f=1/2$) orthogonal ensemble; dashed line:  the
semi-Poisson model; dotted line: orthogonal ensemble.

Fig. 2 (a) form factor; (b) nearest neighbour spacing distribution (NND). Full
line: incomplete ($f=1/2$) continuous picket fence; dotted line:
orthogonal ensemble.

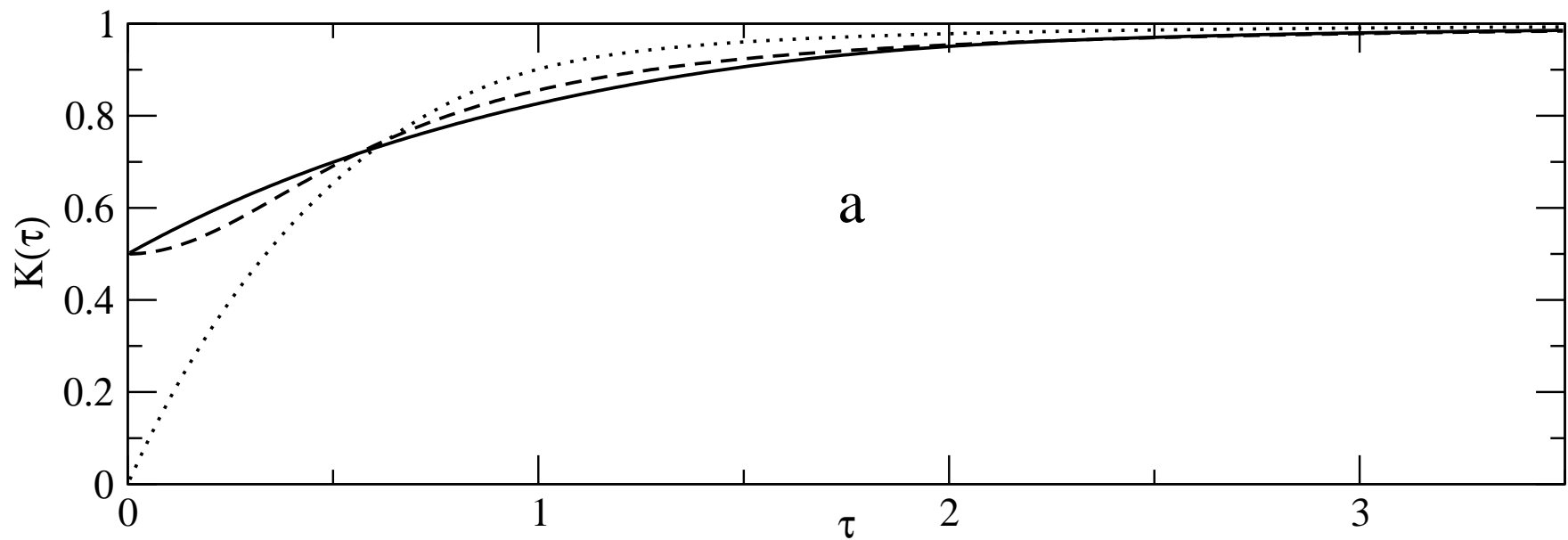
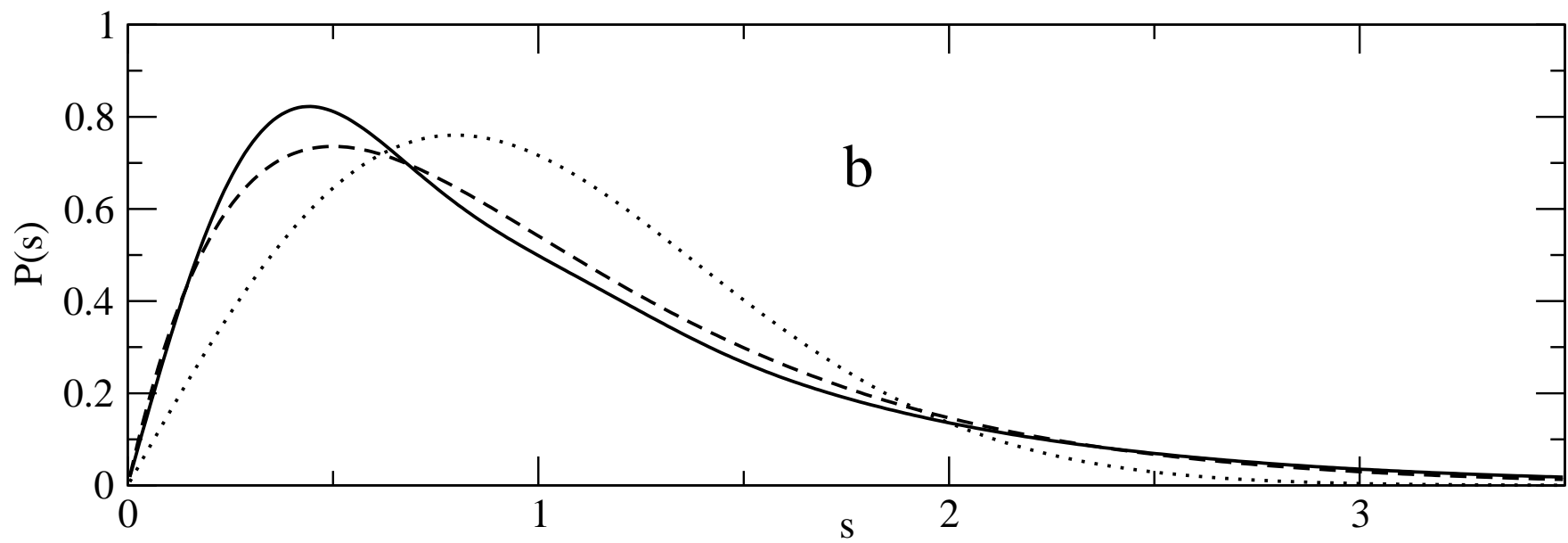
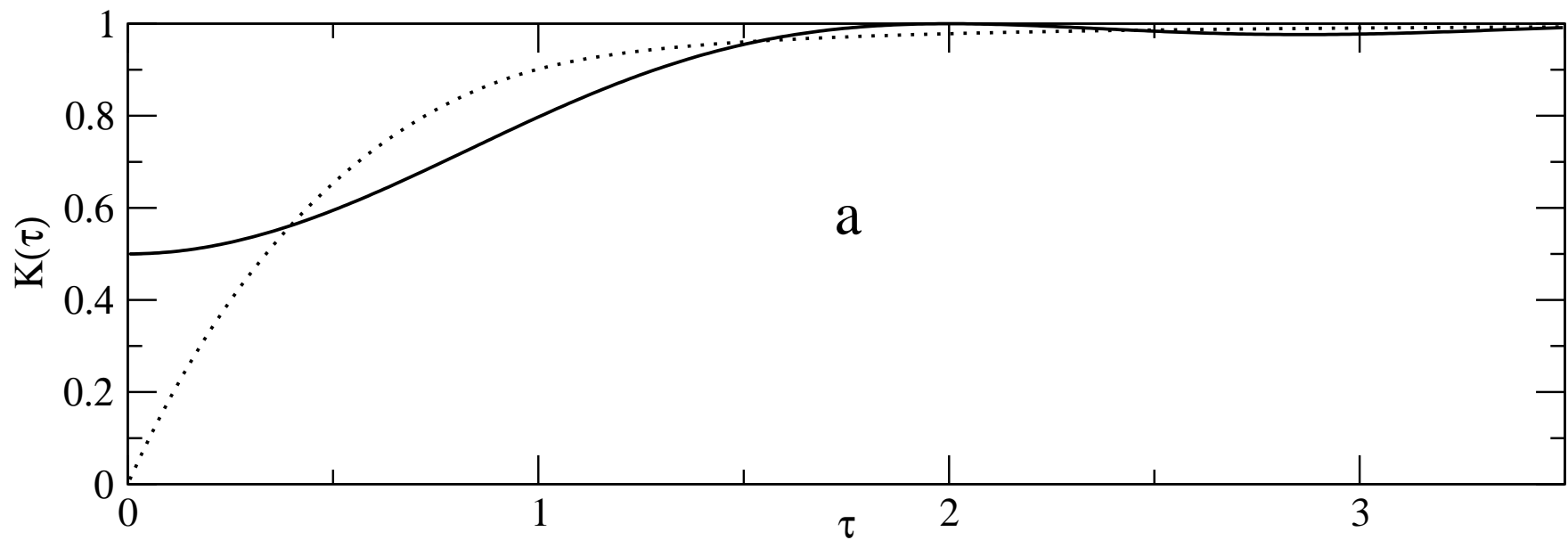
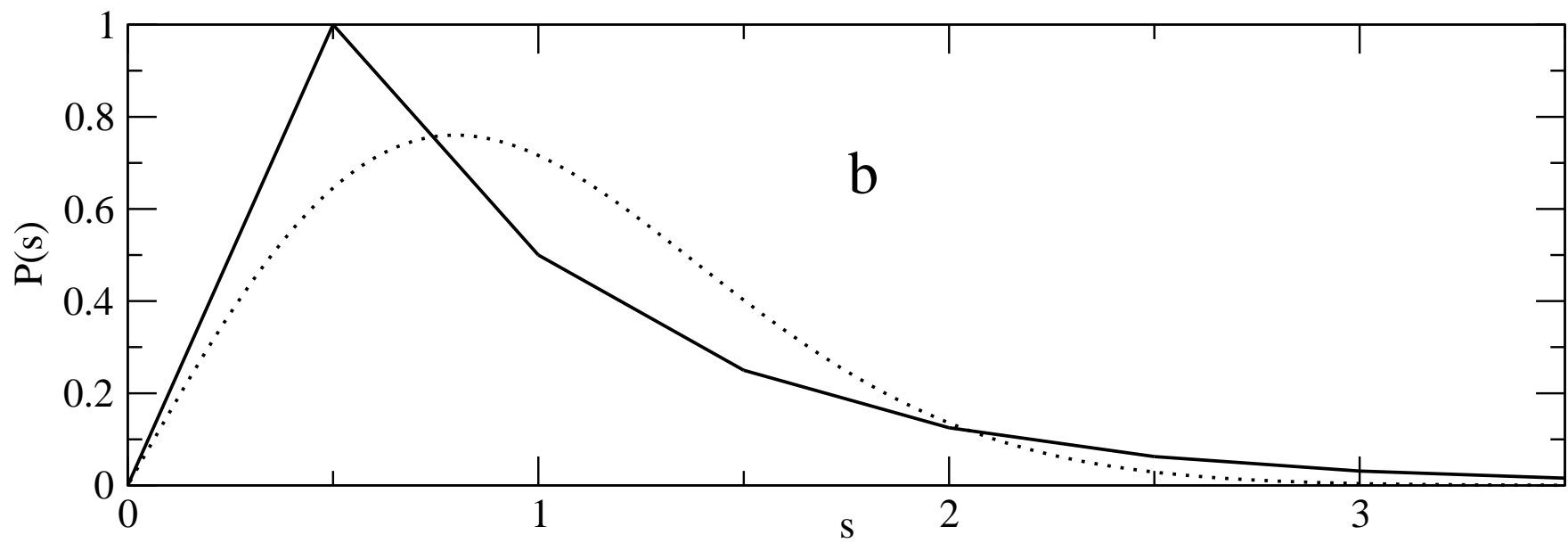
\end{document}